\documentstyle[preprint,aps]{revtex}
\begin{document}

\draft

\title{Re-examination of electronic transports through a quantum wire 
coupled to a quantum dot}

\author{Yu-Liang Liu and T. K. Ng}
\address{Department of Physics, Hong Kong University of Science and 
Technology, Clear Water Bay, Kowloon, Hong Kong, People's Republic of China}

\maketitle

\begin{abstract}

    In this paper we re-examine the problem of electronic transports through
a system consisting of a quantum dot which has well-defined discrete energy 
levels connected to an infinite quantum wire, using the bosonization
method and phase-shift representation, we show that all previously 
known results can be obtained through our method in a very simple way.
Furthermore, the evolution of the system from ultraviolet to infrared 
critical fixed points appears naturally in our method.
\end{abstract}
\vspace{1cm}


\newpage

\section{Introduction}

  Transport properties of a quantum dot connected to two reservoirs (leads)
have been extensively studied\cite{1,2,3,4,5,6,7,8,8'} recently. 
For a large quantum dot where energy levels are approximately continuous, 
the system shows Coulomb blockade behavior due to static Coulomb 
interaction between electrons in the dot, and its conductance is quantized.
The situation is quite different for a very small quantum dot which has 
well-separated energy levels. If the number of electrons in the dot is odd,
which can be controlled by external gate voltages, the system shows
Kondo-like low energy behavior. This property was confirmed
experimentally\cite{9,10} in a system corresponding to quantum dot coupled 
to two leads by tunneling barriers. 

   In this paper, we consider another interesting situation where we 
replace the two reservoirs (leads) by an infinite quantum wire. 
For a very small quantum dot with well-defined energy levels, the 
tunneling conductance is determined by the level which is 
closest to the Fermi level of the quantum wire (assuming other
levels below it are occupied by even electrons and the levels above
are unoccupied). By combining the bosonization method with a phase shift
representation proposed by one of us, we rederive a number
of known results for this problem in this paper, including (i)the zero
temperature $I-V$ characteristic with $I\sim V_{g}^{2/g-1}$ in the two 
limit cases, $\epsilon_{0}>>\Gamma$ and $\epsilon_{0}<<-\Gamma$, where 
$V_{g}$ is an external voltage, $g$ is a dimensionless coupling constant
of conduction electrons, $\Gamma$ is the tunneling width between the
quantum wire and dot, and $\epsilon_{0}$ is the level energy relative 
to the Fermi levels of the two quantum wires, and (ii)in the case 
$\epsilon_{0}\sim 0$, the tunneling current is proportional to the 
external voltage $V_{g}$. Moreover, we demonstrate that even though the
tunneling current has the same low energy behavior in both cases of
$\epsilon_{0}>>\Gamma$ and $\epsilon_{0}<<-\Gamma$, the spin 
susceptibility of the local electron in the quantum dot in the two 
cases are completely different. In the former case where the level
$\epsilon_{0}$ is essentially unoccupied, the local electron level 
only provides a barrier-like potential scattering for conduction
electrons, and its spin susceptibility is zero. In the latter 
case, the level $\epsilon_{0}$ is single-occupied and the local 
electron has a Kondo-like exchange interaction with the conduction 
electrons. Its spin susceptibility shows generally low energy 
power-law behavior. However, for a strongly repulsive 
electron-electron interaction which suppresses the electron density of 
states near the quantum dot due to reflection of the conduction electrons, 
the local spin becomes nearly free, and the spin susceptibility shows 
simple Curie-type behavior.

   In our analysis we shall assume that the quantum dot is embedded in an 
infinite quantum wire, the conduction electrons may tunnel from the quantum 
wire to the quantum dot, or {\it vice versa}. Notice that one may also 
consider another similar system with a quantum dot weakly connected to 
two half-infinite quantum wires with open boundary constraint
conditions\cite{11,12} at the quantum dot site. As shown in
Refs.\cite{15,15a,15',13,14}, in the former case and for electrons 
with repulsive interaction, the system goes to an infrared critical fixed 
point where the electrons are completely reflected from the quantum dot. 
In this limit the two quantum-dot systems should show similar low energy 
behavior of tunneling $I-V$. Notice, however that the low energy physics 
in these two cases are in general different. In the former case the 
complete reflection of electrons on the quantum dot is induced by the 
backward scattering potential which is relevant, while in the latter it is a 
boundary constraint and the backward scattering potential becomes irrelevant.

We shall describe our system by the following Hamiltonian,
\begin{eqnarray}
H &=& H_{0}+H_{d}+H_{t} \nonumber \\
H_{0} &=& \displaystyle{ -i\hbar v_{F} \int dx[\psi^{\dagger}_{R\sigma}(x)
\partial_{x}\psi_{R\sigma}(x)-\psi^{\dagger}_{L\sigma}(x)\partial_{x}
\psi_{L\sigma}(x)] } \nonumber \\
&+& \displaystyle{ \frac{V}{2}\int dx \rho_{R\sigma}(x)\rho_{L\sigma}(x)
+\frac{eV_{g}}{2}\sum_{\sigma}\int dx [\rho_{R\sigma}(x)-\rho_{L\sigma}(x)]}
\label{1} \\
H_{d} &=& \displaystyle{ \epsilon_{0}d^{\dagger}_{\sigma}d_{\sigma}+
Ud^{\dagger}_{\uparrow}d_{\uparrow}d^{\dagger}_{\downarrow}d_{\downarrow}}
\nonumber \\
H_{t} &=& \displaystyle{ t_{0}\{d^{\dagger}_{\sigma}[\psi_{R\sigma}(0)+
\psi_{L\sigma}(0)]+[\psi^{\dagger}_{R\sigma}(0)+\psi^{\dagger}_{L\sigma}(0)]
d_{\sigma} \}} \nonumber
\end{eqnarray}
where $\psi_{R\sigma}(x)$ are the right-moving electron fields, 
$\psi_{L\sigma}(x)$ are the left-moving electron fields, 
$\rho_{R(L)\sigma}(x)=\psi^{\dagger}_{R(L)\sigma}(x)\psi_{R(L)\sigma}(x)$,
$V_{g}$ is an external
voltage added between two ends of the quantum dot, $d_{\sigma}$ describe the 
electron in the level $\epsilon_{0}$ of the quantum dot, $U$ is the on 
site Coulomb repulsion, and $t_{0}$ is the hybridization amplitude
between the quantum dot and wire. Here, we have assumed that the quantum dot 
is small enough to have well separated levels $\Delta \epsilon= \epsilon_{n}-
\epsilon_{n-1} \gg \Gamma$, and the dot can be seen as a structureless 
point at coordinate space. This model cannot be rigorously solved, and 
its physical behavior strongly depends on the parameters $\epsilon_{0}$ 
and $U$. In the low energy region, the tunneling $I-V$ of the system is 
mainly controlled by the occupation number of the level $\epsilon_{0}$
of the quantum dot (assuming other levels below it are occupied by even
electrons and levels above are unoccupied). The Hilbert space of the local
electron state $d_{\sigma}$ composes of four states $\{|0>,\;|\uparrow>,\;
|\downarrow>,\;|\uparrow\downarrow>\}$, 
and its ground state depends on the value of $\epsilon_{0}$.
In section ${I\!\!I}$, shall we consider the case with $<d^{\dagger}_{\sigma}
(t)d_{\sigma}(t)>\cong 0$, i.e., $\epsilon_{0} >> \Gamma$, the level is 
far above the Fermi level of the quantum wire. In this limit, the quantum 
dot appears as a potential barrier to the conduction electrons. Using
bosonization and phase shift representations, we show the evolution of 
the system from ultraviolet to infrared critical fixed points, and recover
the known low energy physics at the infrared critical fixed point.
In section ${ I\!\!I\!\!I}$, we study the case with $0<\; 
<d^{\dagger}_{\sigma}(t)d_{\sigma}(t)>\; <1$, and $|\epsilon_{0}|\leq\Gamma$.
In this case, the conduction electrons resonant with the electron level 
$d_{\sigma}$ in the quantum dot. In section ${I\!\!V}$, we consider the 
case $<d^{\dagger}_{\sigma}(t)d_{\sigma}(t)>\cong 1$, i.e., $\epsilon_{0} 
<< -\Gamma$, the level $\epsilon_{0}$ is far below the Fermi level of the 
quantum wire. In this case, the system is in the usual Kondo regime of 
Luttinger liquid. We shall show that all known previous results on this
problem can be obtained in relatively straightforward ways using our 
approach. Finally, we give our conclusion and discussions in section 
$V$ where the advantage of our method will be pointed out.

\section{($\epsilon_{0}>> \Gamma$) non-magnetic impurity-like scattering 
in the non-equilibrium regime}

In the case of $\epsilon_{0}>> \Gamma$, the level $\epsilon_{0}$ is 
unoccupied at equilibrium. The $U$-term in $H_{d}$ can be neglected, and 
we can simply integrate out the electron field $d_{\sigma}$ to obtain an 
effective Hamiltonian
\begin{eqnarray}
H_{eff.} &=& H_{0}+H_{im} \nonumber \\
H_{im} &=& \displaystyle{ U_{0}[\psi^{\dagger}_{R\sigma}(0)\psi_{R\sigma}(0)+
\psi^{\dagger}_{L\sigma}(0)\psi_{L\sigma}(0)] } \label{2} \\
&+& \displaystyle{ U_{2k_{F}}[\psi^{\dagger}_{R\sigma}(0)\psi_{L\sigma}(0)+
\psi^{\dagger}_{L\sigma}(0)\psi_{R\sigma}(0)]}, \nonumber
\end{eqnarray}
where $U_{0}\sim U_{2k_{F}}\sim t^{2}_{0}/\epsilon_{0}\sim\Gamma$. The
$U_{0}$-term represents usual forward scattering which does not 
influence the transport behavior of the system, and can be neglected. 
The $U_{2k_{F}}$-term describes usual backward scattering which 
determines the low energy behavior of the system, i.e. the system 
reduces to a single impurity scattering problem. In usual equilibrium 
renormalization group treatment with $V_{g}=0$, the backward scattering
potential $U_{2k_{F}}$ is a relevant quantity for repulsive 
electron-electron interaction $V>0$, and has a renormalized
form\cite{15} $U_{2k_{F}}(l)=U_{2k_{F}}e^{(1-g)l}$, where 
$g=(\frac{2\pi\hbar v_{F}-V}{2\pi\hbar v_{F}+V})^{1/2}$, and $l$ is the
renormalization scale parameter. In the low energy limit 
$l\rightarrow \infty$, the renormalized backward scattering potential
$U_{2k_{F}}(l)$ goes to infinity and the low energy behavior of the system 
is determined by an effective (renormalized) Hamiltonian which can be 
obtained from (\ref{2}) with the backward scattering potential $U_{2k_{F}}$ 
replaced by the renormalized one $U_{2k_{F}}(l)$. Due to the singular low 
energy behavior of the backward scattering potential, it is inconvenient 
to study the low energy behavior of the system with this effective 
Hamiltonian. One has to conjecture\cite{15} that at zero temperature where 
the backward scattering potential $U_{2k_{F}}(l)$ goes to infinity, the
electrons are completely reflected by the quantum dot at $x=0$. This 
conjecture can be justified with an alternative phase representation 
of the model\cite{13,14} which we shall discuss in the following.

   Here we consider this problem in the non-equilibrium case\cite{14'} 
$V_{g}\neq 0$. For simplicity, we consider spinless fermions because at 
present the spin freedom appears only as a channel index. We first 
introduce a set of new electron fields
\begin{equation}
\psi_{1}(x)=\frac{1}{\sqrt{2}}(\psi_{R}(x)+\psi_{L}(-x)), \;\;\;
\psi_{2}(x)=\frac{1}{\sqrt{2}}(\psi_{R}(x)-\psi_{L}(-x)).
\label{3}\end{equation}
It is easy to check that the operators $\psi_{1(2)}(x)$ obey the standard 
anticommutation relations. The bosonic representation\cite{16,17,18} of
$\psi_{1(2)}(x)$ is $\psi_{1(2)}(x)=(D/2\pi)^{1/2}\exp\{-i\Phi_{1(2)}(x)\}$,
where $\partial_{x}\Phi_{1(2)}(x)=2\pi\rho_{1(2)}(x)$.
In terms of these new fields, the Hamiltonian
(\ref{2}) can be written as
\begin{eqnarray}
H_{eff.} &=& H_{0}+H_{im} \nonumber \\
H_{0} &=& \displaystyle{-i\int dx[\psi^{\dagger}_{1}(x)\partial_{x}\psi_{1}(x)
+\psi^{\dagger}_{2}(x)\partial_{x}\psi_{2}(x)]} \nonumber \\
&+& \displaystyle{\frac{V}{4}\int dx \{ [\rho_{1}(x)+\rho_{2}(x)]
[\rho_{1}(-x)+\rho_{2}(-x)]} \nonumber \\
&-& \displaystyle{ [\psi^{\dagger}_{1}(x)\psi_{2}(x)+h.c.]
[\psi^{\dagger}_{1}(-x)\psi_{2}(-x)+h.c.]} \label{4} \\
&+& \displaystyle{\frac{eV_{g}}{2}\int dx [\psi^{\dagger}_{1}(x)\psi_{2}(x)
+\psi^{\dagger}_{2}(x)\psi_{1}(x)]} \nonumber \\
H_{im} &=& \displaystyle{U_{2k_{F}} [\rho_{1}(0)-\rho_{2}(0)]}
\nonumber\end{eqnarray}
where $\rho_{1(2)}(x)=\psi^{\dagger}_{1(2)}(x)\psi_{1(2)}(x)$. We have 
neglected the forward scattering term which doses not influence the transport
properties of the system.

In the phase shift representation\cite{13,14}, we obtain an effective 
renormalized Hamiltonian directly from (\ref{4})
\begin{eqnarray}
H_{eff.} &=& H_{0}+H_{IM} \nonumber \\
H_{IM} &=& \displaystyle{ \hbar v_{F}\delta[\rho_{1}(0)-\rho_{2}(0)] }
\label{5}\end{eqnarray}
which describes the low energy behavior of the system with original Hamiltonian
(\ref{2}) or (\ref{4}), where the phase shift $\delta$ is a renormalized 
quantity, $\delta=\arctan[e^{(\frac{1}{g}-1)l}\tan(\delta_{0})]$,
$\delta_{0}=U_{2k_{F}}/(\hbar v_{F})$. In Ref.\cite{14} we derive a
renormalization group equation for $\delta$ for small electron-electron
interaction $V/(2\pi\hbar v_{F})\ll 1$, and the renormalized phase shift
is $\delta\sim\arctan[e^{\gamma l}\tan(\delta_{0})]$, where $\gamma=V/(2\pi
\hbar v_{F})$. However, we have shown in Refs.\cite{13,14} that the
dimensionless coupling constant $g$ determines all correlation functions'
exponents. We therefore replace $\gamma$ by $1/g-1$ since $1/g-1\sim\gamma$ 
at small $\gamma$. At zero temperature and $V_{g}=0$, the phase shift 
$\delta$ take a critical value $\delta^{c}=\pi/2$ at $l\rightarrow\infty$, 
which corresponds to the $U_{2k_F}(l)\rightarrow\infty$ infrared critical 
point of the system. Notice that the backward scattering term $H_{IM}$ 
is {\em finite} at this infrared critical point in the phase shift
representation. Therefore, the effective Hamiltonian (\ref{5}) can be 
used safely to study the transport properties and evolution of 
correlation exponents of the system from high energy ($l=0$) to low 
energy ($l\rightarrow\infty$) regions. 

  It is convenient to introduce the unitary transformation 
$\bar{H}_{eff.}=\hat{U}^{\dagger}(H_{0}+H_{IM})\hat{U}$, where
\begin{equation}
\hat{U}=\exp\{i\frac{\delta}{2\pi}[\Phi_{1}(0)-\Phi_{2}(0)]\},
\label{6}\end{equation}
and $\partial_{x}\Phi_{1(2)}(x)=2\pi\rho_{1(2)}(x)$.
Together with the global gauge transformations
\begin{equation}
\psi_{1(2)}(x)=\bar{\psi}_{1(2)}(x)e^{i\theta_{1(2)}}, \;\;\;\;
\theta_{1}-\theta_{2}=\delta,
\label{7}\end{equation}
we obtain the effective Hamiltonian $\bar{H}^{c}_{eff.}$ at the infrared 
critical point 
($\delta^{c}=\pi/2$)\cite{13,14} 
\begin{eqnarray}
\bar{H}^{c}_{eff.} &=& \displaystyle{-i\hbar v_{F}\int dx 
[\bar{\psi}^{\dagger}_{R}(x)\partial_{x}
\bar{\psi}_{R}(x)-\bar{\psi}^{\dagger}_{L}(x)\partial_{x}\bar{\psi}_{L}(x)]}
\nonumber \\
&+& \displaystyle{\frac{V}{2}\int dx[\bar{\rho}_{R}(x)\bar{\rho}_{R}(-x)
+\bar{\rho}_{L}(x)\bar{\rho}_{L}(-x)]}
\label{8} \\
&-& \displaystyle{\frac{eV_{g}}{2}\int^{\infty}_{0}dx[
\bar{\rho}_{R}(x)-\bar{\rho}_{R}(-x)+\bar{\rho}_{L}(x)-\bar{\rho}_{L}(-x)]}
\nonumber\end{eqnarray}
where $\bar{\psi}_{R}(x)=[\bar{\psi}_{1}(x)+\bar{\psi}_{2}(x)]/\sqrt{2}$,
$\bar{\psi}_{L}(-x)=[\bar{\psi}_{1}(x)-\bar{\psi}_{2}(x)]/\sqrt{2}$; 
$\bar{\rho}_{R(L)}(x)=
\bar{\psi}^{\dagger}_{R(L)}(x)\bar{\psi}_{R(L)}(x)$. 
This Hamiltonian $\bar{H}^{c}_{eff.}$ completely determines the low energy
physical behavior of the system. Note that at the infrared critical 
point $\delta^{c}=\pi/2$, the effect of the backward scattering induced 
by the quantum dot is to alter the interaction among the conduction 
electrons, and to completely separate the right- and left-moving 
electrons. It also alters the coupling to the external voltage due to 
reflection of the electrons from the quantum dot. 

The reflection of the conduction electrons on the quantum dot can be seen
clearly from the relations
\begin{eqnarray}
\hat{U}^{\dagger}\psi_{R}(x)\hat{U} &=& \displaystyle{e^{i\theta_{1}}
e^{-i\delta/2}
\cdot\left\{ \begin{array}{ll}
\cos(\delta)\bar{\psi}_{R}(x)+i\sin(\delta)\bar{\psi}_{L}(-x), &\;\;\;\; x>0 \\
\bar{\psi}_{R}(x), & \;\;\;\; x<0 \end{array}\right.} \nonumber \\
\hat{U}^{\dagger}\psi_{L}(x)\hat{U} &=& \displaystyle{e^{i\theta_{1}}
e^{-i\delta/2} \cdot\left\{ \begin{array}{ll}
\bar{\psi}_{L}(x), & \;\;\;\; x>0 \\
\cos(\delta)\bar{\psi}_{L}(x)+i\sin(\delta)\bar{\psi}_{R}(-x). &\;\;\;\; x<0
\end{array}\right.}
\label{9}\end{eqnarray}
With these relations, we can obtain the reflection and transmission
rates of the electrons, ${\cal R}$ and ${\cal T}$, respectively, where
\begin{equation}
{\cal R}=\sin^{2}(\delta), \;\;\;\; {\cal T}=\cos^{2}(\delta).
\label{10}\end{equation}
We shall show in the following that the tunneling conductance is 
proportional to the transmission rate ${\cal T}$. At the infrared 
critical point $\delta^{c}=\pi/2$, equation (\ref{9}) shows that the 
right- and left-moving electrons are completely reflected on
the quantum dot,
\begin{eqnarray}
\hat{U}^{\dagger}\psi_{R}(x)\hat{U}|_{\delta=\delta^{c}} &=& \displaystyle{
e^{i\theta_{1}-i\pi/4}\cdot \left\{ \begin{array}{ll}
i\bar{\psi}_{L}(-x), & \;\;\;\; x>0 \\
\bar{\psi}_{R}(x), & \;\;\;\; x<0 \end{array}\right.} \nonumber \\
\hat{U}^{\dagger}\psi_{L}(x)\hat{U}|_{\delta=\delta^{c}} &=& \displaystyle{
e^{i\theta_{1}-i\pi/4}\cdot \left\{ \begin{array}{ll}
\bar{\psi}_{L}(x), & \;\;\;\; x>0 \\
i\bar{\psi}_{R}(-x). & \;\;\;\; x<0 \end{array}\right.}
\nonumber\end{eqnarray}
The external voltage term can be written in a more compact form by taking 
another transformation,
\begin{eqnarray}
\bar{\psi}_{R}(x) &\rightarrow & \displaystyle{ \left\{ \begin{array}{ll}
e^{ieV_{g}x/(\hbar v_{F})}\bar{\psi}_{R}(x), & \;\;\;\; x>0 \\
\bar{\psi}_{R}(x), & \;\;\;\; x<0 \end{array}\right.} \nonumber \\
\bar{\psi}_{L}(x) &\rightarrow & \displaystyle{ \left\{ \begin{array}{ll}
\bar{\psi}_{L}(x), & \;\;\;\; x>0 \\
e^{ieV_{g}x/(\hbar v_{F})}\bar{\psi}_{L}(x). & \;\;\;\; x<0 \end{array}
\right.}\label{11}\end{eqnarray}
With this the effective Hamiltonian (\ref{8}) can be rewritten as
\begin{eqnarray}
\bar{H}^{c}_{eff.} &=& \displaystyle{ -i\hbar v_{F}\int dx[
\bar{\psi}^{\dagger}_{R}(x)\partial_{x}\bar{\psi}_{R}(x)-
\bar{\psi}^{\dagger}_{L}(x)\partial_{x}\bar{\psi}_{L}(x)]} \nonumber \\
&+& \displaystyle{ \frac{V}{2}\int dx[\bar{\rho}_{R}(x)\bar{\rho}_{R}(-x)+
\bar{\rho}_{L}(x)\bar{\rho}_{L}(-x)]} \label{12} \\
&+& \displaystyle{ \frac{eV_{g}}{2}\int dx[\bar{\rho}_{R}(x)-
\bar{\rho}_{L}(x)]}. \nonumber
\end{eqnarray}

To study low-energy transports we first employ the equations 
$e\partial_{t}[\rho_{R}(x,t)+\rho_{L}(x,t)]+\partial_{x}\hat{J}(x,t)=0$, 
and $\partial_{t}\rho_{R(L)}(x,t)=\frac{i}{\hbar}
[H_{eff.}, \rho_{R(L)}(x,t)]$ to obtain the current operator
\begin{equation}
\hat{J}(x,t)=ev_{F}(1-\gamma)[\rho_{R}(x,t)-\rho_{L}(x,t)]
\label{13}\end{equation}
where $\gamma=V/(2\pi\hbar v_{F})$. We have used the bosonized 
expression\cite{16,17,18} of
the Hamiltonian $H_{eff.}$ to obtain the current operator $\hat{J}(x,t)$.
The factor $\gamma$ presents the influence of the electron-electron
interaction on the current. Under the unitary transformation $\hat{U}$, 
the current operator becomes
\begin{eqnarray}
\hat{\bar{J}}(x,t) &=& \hat{U}^{\dagger}\hat{J}(x,t)\hat{U} \nonumber \\
&=& \displaystyle{ ev_{F}(1-\gamma)\cdot\left\{\begin{array}{ll}
\alpha\bar{\rho}_{R}(x,t)+\beta\bar{\rho}_{L}(-x,t)-\bar{\rho}_{L}(x,t)-
\hat{I}^{'}(x,t), & \;\;\;\; x>0 \\
\bar{\rho}_{R}(x,t)-\beta\bar{\rho}_{R}(-x,t)-\alpha\bar{\rho}_{L}(x,t)-
\hat{I}^{'}(-x,t), & \;\;\;\; x<0 \end{array}\right.}
\label{14} \\
\hat{I}^{'}(x,t) &=& \displaystyle{ i\sin(2\delta)[e^{i2eV_{g}/(\hbar v_{F})} 
\bar{\psi}^{\dagger}_{L}(-x,t)\bar{\psi}_{R}(x,t)-
e^{-i2eV_{g}/(\hbar v_{F})}
\bar{\psi}^{\dagger}_{R}(x,t)\bar{\psi}_{L}(-x,t)]}
\nonumber\end{eqnarray}
where $\alpha=[1+\cos(2\delta)]/2$, and $\beta=[1-\cos(2\delta)]/2$.
With this expression, it is straightforward to obtain the tunneling current
\begin{eqnarray}
\bar{I} &=& <\hat{\bar{J}}(x,t)> \nonumber \\
&=& \frac{ge^{2}}{2\pi\hbar}V_{g}{\cal T},
\label{15}\end{eqnarray}
and with corresponding tunneling conductance, 
$G=\frac{ge^{2}}{2\pi\hbar}{\cal T}$. At zero temperature and with small 
finite external voltage $V_{g}$ which is taken as a low energy cutoff 
factor, the renormalization parameter saturates at $l=-\ln(eV_{g}/D)$, 
where $D$ is the conduction bandwidth, and the renormalized phase shift is
$\delta=\arctan[(eV_{g}/D)^{1-1/g}\tan(\delta_{0})]$, therefore the 
transmission rate ${\cal T}$ of the conduction electrons is
\begin{equation}
{\cal T}=\frac{(eV_{g})^{2(\frac{1}{g}-1)}}{
(eV_{g})^{2(\frac{1}{g}-1)}+D^{2(\frac{1}{g}-1)}\tan^{2}(\delta_{0})}.
\label{16}\end{equation}
In the case $\delta_{0}=0$ (no quantum dot), the tunneling current 
(\ref{15}) retores the usual form of an infinite quantum wire. For
$\delta_{0}\neq 0$, the tunneling current (\ref{15}) is also consistent 
with results from previous calculations\cite{12,15,19}. 

The simple expression of the transmission rate is one of our central 
results, and is exact to leading order in $V_g$. For the special 
case of $g=1/2$, this model can be exactly solved, and the exact 
expression of the tunneling current is known\cite{15}. As expected,
difference between the tunneling current (\ref{15}) and the exact 
result appears only in higher order. The deviation comes from two 
sources, one is from the renormalized phase shift which is obtained 
by perturbation method in our analysis, and the other one comes from 
our approximation of using only the critical point Hamiltonian
$\bar{H}^{c}_{eff.}$ to calculate the tunneling current. Notice that 
in Ref.\cite{19a}, the authors calculated the tunneling current for 
a weak interaction ($g\sim 1$) electron system in Born approximation, 
and their result is consistent with the tunneling current (\ref{15}) 
in the small $V$ limit $1/g\sim 1+V/(2\pi\hbar v_{F})$. In 
Ref.\cite{guinea} the authors argued that the frequency dependence of
conductance should include two terms: one is $c_{1}\omega^{2}$, and 
another one is $c_{2}\omega^{2/g-2}$, where $\omega$ is frequency, $c_{1}$
and $c_{2}$ are constants. The $\omega^{2}$-term is not universal, and
depends on an artificial finite cut-off, such as $\omega_{c}$ in Eq.(5) of  
Ref.\cite{guinea}. For a finite quantum wire, one can choose 
$c_1\sim\hbar v_{F}q_{c}$ as a reasonable cut-off, where $q_{c}=2\pi/L$, 
$L$ is the length of the system. For an infinite quantum wire, $c_{1}$
goes to zero and the $\omega^{2}$-term does not appear.

We have so far used only the effective Hamiltonian at the infrared 
critical point $\delta^{c}=\pi/2$ to calculate the tunneling current. 
However, rigorously speaking, in the presence of small finite external 
voltage $V_g$, the system will deviate from the infrared critical 
point, and the low energy physical properties of the system would be 
determined by the Hamiltonian $H_{T}=\bar{H}^{c}_{eff.}+\Delta H$, where 
$\Delta{H}=-\hbar{v}_{F}(\delta^{c}-\delta)[\bar{\psi}^{\dagger}_{R}(0)
\bar{\psi}_{L}(0)+\bar{\psi}^{\dagger}_{L}(0)\bar{\psi}_{R}(0)]$ is a 
small perturbation around the infrared critical point. The quantity 
$\delta^{c}-\delta\ll 1$ presents the deviation of the system away from 
the infrared critical point. We shall now show that the perturbative 
term $\Delta H$ only contributes to high order correction to the 
tunneling current. The corrections to tunneling current from $\Delta H$ 
can be estimated perturbatively. We obtain,
\begin{eqnarray}
\bar{I} &=& \displaystyle{ <\hat{\bar{J}}(x,t)> -\frac{i}{L}\int dx\int dt^{'}
<\hat{\bar{J}}(x,t)\Delta H(t^{'})> } \nonumber \\
&-& \displaystyle{ \frac{1}{L}\int dx\int dt^{'}dt^{``}<\hat{\bar{J}}(x,t)
\Delta H(t^{'})\Delta H(t^{``})> +\; ...}
\label{17}\end{eqnarray}
where we have taken an average over space coordinate $x$, and $L\rightarrow
\infty$ is the length of the system. The leading order correction comes from
the operator $\hat{I}^{'}(x,t)$, and is proportional to $(eV_{g})^{2(1/g-1)}
(\delta^{c}-\delta)\sin(2\delta)/L$, where $\delta^{c}-\delta$ can be written
as $\delta^{c}-\delta=\arcsin[\cos(\delta)]$. For small external voltage
$V_{g}$, this correction can be neglected safely.

   Before proceeding to next section, we first summarize our findings 
so far. Using the phase shift representation, we have calculated the 
tunneling current for the quantum-dot problem in the limit $\epsilon_0>>
\Gamma$, in which the problem can be mapped to the problem of \
non-magnetic impurity scattering. We obtain results in agreement with
previous perturbation\cite{12,15} and Bethe ansatz\cite{19} calculations.
The phase representation method has the advantage that explicit divergence 
in the effective backward scattering amplitude is avoided, and the 
influence of the backward scattering on the system, such as the evolution 
of the Hamiltonian from high energy to low energy regions, the change of 
the effective interaction among the electrons, and the evolution of 
correlation exponents from ultraviolet ($\delta=0$) to infrared 
($\delta^{c}=\pi/2$) critical points, can be seen clearly. By introducing
a simple unitary transformation on the original electron fields, we can 
obtain in a straightforward way the reflectivity and transmission rates 
of the electrons through the quantum dot, and then obtain the tunneling 
conductance of the system. It is one of the most prominent character of
one-dimensional interacting electron systems that the backward scattering 
of electrons on impurity (barrier) alters the effective electron-electron 
interaction because the right- and left-moving electrons are mixed by this
scattering. This change in effective electron-electron interaction induces 
the observed singular low energy behaviors in the system. In the 
phase representation this line of physics is demonstrated clearly and
naturally.

\section{($\epsilon_{0}\sim 0$) Electron resonant scattering 
by the local level}

In the case $\epsilon_{0}\sim 0$, the level $\epsilon_{0}$ is close
to the Fermi level of the quantum wire, and its electron occupation number 
satisfies $0<\;<d^{\dagger}_{\sigma}(t)d_{\sigma}(t)>\;<1$. The resonance
between the conduction electrons $\psi_{R(L)\sigma}(x)$ and the local 
electron $d_{\sigma}$ determines the low energy properties of the 
system. We shall treat the term $Ud^{\dagger}_{\uparrow}d_{\uparrow}
d^{\dagger}_{\downarrow}d_{\downarrow}$ by mean field theory in the 
following. In this approximation the $U$ term only renormalize the
energy\cite{19'} $\epsilon_{0}$ and the Hamiltonian (\ref{1}) reduces to
\begin{eqnarray}
H &=& \displaystyle{ -i\hbar v_{F}\int dx[\psi^{\dagger}_{R\sigma}(x)
\partial_{x}\psi_{R\sigma}(x)-\psi^{\dagger}_{L\sigma}(x)\partial_{x}
\psi_{L\sigma}(x)]} \nonumber \\
&+& \displaystyle{ V\int dx \rho_{R\sigma}(x)\rho_{L\sigma}(x)+\frac{eV_{g}}
{2}\sum_{\sigma}\int dx[\rho_{R\sigma}(x)-\rho_{L\sigma}(x)]} \nonumber \\
&+& \displaystyle{ t_{0}\{[\psi^{\dagger}_{R\sigma}(0)+
\psi^{\dagger}_{L\sigma}(0)]d_{\sigma}+d^{\dagger}_{\sigma}[\psi_{R\sigma}(0)+
\psi_{L\sigma}(0)]\}} \label{18} \\
&+& \displaystyle{ U_{2k_{F}}[\psi^{\dagger}_{R\sigma}(0)\psi_{L\sigma}(0)
+\psi^{\dagger}_{L\sigma}(0)\psi_{R\sigma}(0)]+\tilde{\epsilon}_{0}
d^{\dagger}_{\sigma}d_{\sigma}}
\nonumber\end{eqnarray}
where $\tilde{\epsilon}_{0}$ is the modified energy of the local electron
$d_{\sigma}$. The Hamiltonian (\ref{18}) is different from the previous
one in section II (\ref{2}) by the presence of on site energy term for 
the local electron and the hybridization term between the conduction 
electrons and the local electron. The bare backward scattering 
potential $U_{2k_{F}}$ is small because of strong resonance between 
the conduction electrons and the local electron orbital $d_{\sigma}$. 
However, the backward scattering term is relevant, no matter how small 
the bare backward scattering potential is, and will be renormalized to 
infinity in the low energy limit. The low energy behavior of the system is
thus controlled by an effective Hamiltonian in which the renormalized 
backward scattering potential goes to infinity at the infra-red critical
point and the main transport comes from resonant tunneling. Notice that
due to absence of spin exchange interactions in (\ref{18}), the spin 
freedom appears only as a channel index. For simplicity, we shall only 
consider a spinless model with the same interaction form as the Hamiltonian
(\ref{18}). Using the same method as that in Sect.${I\!\!I}$, we obtain
the effective Hamiltonian at the infrared critical point $\delta^{c}=\pi/2$,
\begin{eqnarray}
\bar{H}^{c} &=& \displaystyle{ -i\hbar v_{F}\int dx[\bar{\psi}^{\dagger}_{R}
(x)\partial_{x}\bar{\psi}_{R}(x)-\bar{\psi}^{\dagger}_{L}(x)\partial_{x}
\bar{\psi}_{L}(x)] } \nonumber \\
&+& \displaystyle{ \frac{V}{2}\int dx[\bar{\rho}_{R}(x)\bar{\rho}_{R}(-x)+
\bar{\rho}_{L}(x)\bar{\rho}_{L}(-x)]} \label{19} \\
&+& \displaystyle{ \frac{eV_{g}}{2}\int dx [\bar{\rho}_{R}(x)-
\bar{\rho}_{L}(x)]} \nonumber \\
&+& \displaystyle{ t_{0}\{e^{-i\theta_{1}+i\pi/4}[\bar{\psi}^{\dagger}_{R}(0)
+\bar{\psi}^{\dagger}_{L}(0)]f+e^{i\theta_{1}-i\pi/4}f^{\dagger}[
\bar{\psi}_{R}(0)+\bar{\psi}_{L}(0)] }
\nonumber\end{eqnarray}
where we have replaced $d_{\sigma}$ by $f$ for the spinless case.

Using the continuity equation, we obtain the current operator
\begin{eqnarray}
\hat{J}_{T}(x,t) &=& \displaystyle{ \hat{J}(x,t)+\hat{J}_{r}(t)} \nonumber\\
\hat{J}_{r}(t) &=& \displaystyle{ \frac{iet_{0}}{\hbar}\{[
\psi^{\dagger}_{R}(0,t)-\psi^{\dagger}_{L}(0,t)]f(t)-f^{\dagger}(t)[
\psi_{R}(0,t)-\psi_{L}(0,t)]\}}
\label{20}\end{eqnarray}
where $\hat{J}(x,t)$ is given by Eq.(\ref{13}). $\hat{J}_{r}(t)$ is a 
resonant current operator induced by the resonant term in (\ref{18}).
Under the unitary transformation (\ref{6}), the current operator 
becomes
\begin{eqnarray}
\hat{\bar{J}}_{T}(x,t) &=& \hat{U}^{\dagger}\hat{J}_{T}(x,t)\hat{U} 
\nonumber \\
&=& \hat{\bar{J}}(x,t) +\hat{\bar{J}}_{r}(t) \label{21} \\
\hat{\bar{J}}_{r}(t) &=& \displaystyle{ \frac{iet_{0}}{\hbar}\{
e^{-i\theta_{1}+i\pi/4}[\bar{\psi}^{\dagger}_{R}(0,t)-
\bar{\psi}^{\dagger}_{L}(0,t)]f(t)-e^{i\theta_{1}-i\pi/4}
f^{\dagger}(t)[\bar{\psi}_{R}(0,t)-\bar{\psi}_{L}(0,t)]}
\nonumber\end{eqnarray}
where the current operator $\hat{\bar{J}}(x,t)$ is given by Eq.(\ref{14}).

   Using the effective Hamiltonian (\ref{19}) and the current operator
(\ref{21}), we can calculate the current of the system. The tunneling current
$<\hat{\bar{J}}(x,t)>$ induced by the backward scattering retains the 
same form as that in (\ref{15}) because the resonant term
in (\ref{19}) only contributes a high order correction to it. The resonant
current $<\hat{\bar{J}}_{r}(t)>$ originating from resonant tunneling 
between the conduction electrons and the local electron $f$ can be 
written as
\begin{eqnarray}
\bar{I}_{r} &=& <\hat{\bar{J}}_{r}(t)> \nonumber \\
&=& \displaystyle{ \frac{et_{0}}{\hbar}[G_{f\psi}(t,t)- G_{\psi f}(t,t)]}
\label{22}\end{eqnarray}
where $G_{f\psi}(t,t')=i<[\bar{\psi}^{\dagger}_{R}(0,t)-
\bar{\psi}^{\dagger}_{L}(0,t)]f(t')>e^{-i\theta_{1}+i\pi/4}$, and
$G_{\psi f}(t,t')=i<f^{\dagger}(t)[\bar{\psi}_{R}(0,t')-
\bar{\psi}_{L}(0,t')]>e^{i\theta_{1}-i\pi/4}$. Applying usual 
perturbation method, we obtain for $G_{f\psi}(t,t')$\cite{4,5,7}
\begin{equation}
G_{f\psi}(t,t')=t_{0}\int dt_{1}[G_{RL}(t,t_{1})G^{a}_{f}(t_{1},t')+
G^{r}_{RL}(t,t_{1})G_{f}(t_{1},t')]
\label{23}\end{equation}
where $G_{RL}(t,t')=G_{R}(t,t')-G_{L}(t,t')$, $G_{R(L)}(t,t')=
i<\bar{\psi}^{\dagger}_{R(L)}(0,t)\bar{\psi}_{R(L)}(0,t')>$, $G_{f}(t,t')
=i<f^{\dagger}(t)f(t')>$, and the up index $r$ and $a$  present the
advanced and retarded Green functions, respectively.  
The Green functions satisfy the Dyson equations 
$(G^{a,r}_{f})^{-1}=(G^{0,a,r}_{f})^{-1}-
\sum^{a,r}_{f}$, and $G_{f}=G^{r}_{f}\sum_{f}G^{a}_{f}$, where $\sum^{r,a}_{f}
=t^{2}_{0}[G^{r,a}_{R}+G^{r,a}_{L}]$ and $\sum_{f}=t^{2}_{0}[G_{R}+G_{L}]$.
In the low energy limit, the self-energy has the asymptotic behavior
$\sum^{r,a}_{f}(\omega)\sim \omega^{-1+1/g}$, and the electron Green functions
can be written as\cite{20} $G_{R}(\omega)\sim n(\omega+eV_{g}/2)
\omega^{-1+1/g}$, and $G_{L}(\omega)\sim n(\omega-eV_{g}/2)\omega^{-1+1/g}$, 
where $n(\omega\pm eV_{g}/2)$ is the Fermi-Dirac distribution function. 
Therefore the Green function $G^{r,a}_{f}$ has low-energy asymtotic form
$G^{r,a}_{f}(\omega)\sim\frac{1}{\omega-\epsilon^{'}_{0}+
ia\omega^{-1+1/g}}$, where $\epsilon^{'}_{0}$ is the renormalized energy 
of the local electron $f$ and $a$ is a constant $\sim{t}^{2}_{0}$. To
leading order in $V_g$, the resonant current is given by 
\begin{equation}
\bar{I}_{r}=\frac{ge^{2}}{2\pi\hbar}\frac{a^{2}V^{2/g-1}_{g}}{(eV_{g}-
\epsilon^{'}_{0})^{2}+a^{2}V^{-2+2/g}_{g}}.
\label{24}\end{equation}
Notice that at $eV_{g}=\epsilon^{'}_{0}$, the resonant current 
becomes\cite{15a}
$\bar{I}_{r}=ge^{2}V_{g}/(2\pi\hbar)$, with a linear relation 
between the current and the external voltage, similar to the case of 
free electron system. In the limit $\epsilon^{'}_{0}\gg eV_{g}$, we 
obtain\cite{15a} $\bar{I}_{r}\propto V^{2/g-1}_{g}/\epsilon^{'2}_{0}$. In the 
former case, the current of the system is dominant by the resonant 
current. In the latter case, the resonant current only contributes a small
correction to the current, which is consistent with our calculation in 
Section ${I\!\!I}$ where we consider the limit of large$\epsilon_{0}$
and the local electron field $d_{\sigma}$ is integrated out. The resonant 
current (\ref{24}) provides an interpolation between these two regions which
can be tested experimentally.

\section{($\epsilon_{0}<< -\Gamma$) Quantum scattering of 
a spin-1/2 magnetic impurity}

In the case of $\epsilon_{0}<<-\Gamma$, the level $\epsilon_{0}$ is always
occupied by an electron at equilibrium, $<d^{\dagger}_{\sigma}(t)
d_{\sigma}(t)>\sim1$. In the large $U$ limit, and after making the
Schrieffer-Woelf transformation, we obtain an effective Hamiltonian 
which describes the system in the Kondo regime,
\begin{eqnarray}
H &=& H_{0}+H_{K} \nonumber \\
H_{0} &=& \displaystyle{ -i\hbar v_{F}\int dx[\psi^{\dagger}_{R\sigma}(x)
\partial_{x}\psi_{R\sigma}(x)-\psi^{\dagger}_{L\sigma}(x)\partial_{x}
\psi_{L\sigma}(x)]} \nonumber \\
&+& \displaystyle{ V\int dx\rho_{R\sigma}(x)\rho_{L\sigma}(x)+\frac{eV_{g}}
{2}\sum_{\sigma}\int dx[\rho_{R\sigma}(x)-\rho_{L\sigma}(x)]} \label{25} \\
H_{K} &=& \displaystyle{
\sum_{j}J^{j}_{1}[s_{Rj}(0)+s_{Lj}(0)]\cdot S_{j}+
\sum_{j}J^{j}_{2}[s_{RLj}(0)+s_{LRj}(0)]\cdot S_{j}}
\nonumber\end{eqnarray}
where ${\bf S}=\frac{1}{2}d^{\dagger}_{\alpha}{\bf \sigma}_{\alpha\beta}
d_{\beta}$ is a local spin operator,
${\bf s}_{R(L)}(0)=\frac{1}{2}\psi^{\dagger}_{R(L)\alpha}(0)
{\bf \sigma}_{\alpha\beta}\psi_{R(L)\beta}(0)$, ${\bf s}_{RL(LR)}(0)=
\frac{1}{2}\psi^{\dagger}_{R(L)\alpha}(0){\bf \sigma}_{\alpha\beta}
\psi_{L(R)\beta}(0)$, and $\left(J^{j}_{1},\; J^{j}_{2},\; j=x,y,z
\right)\sim t^{2}_{0}/U$.
The local spin operator satisfies $<S^{2}_{z}(t)>=1/4$, and its Hilbert 
space consists of two states $\{ |\uparrow>,\; |\downarrow>\}$.
In abelian bosonization the term $J^{z}_{2}[s_{RLz}(0)+s_{LRz}(0)]S_{z}$ 
in (\ref{25}) acts as a backward scattering term which is relevant, 
whereas the term $\sum_{j}J^{j}_{1}[s_{Rj}(0)+s_{Lj}(0)]
\cdot S_{j}$ is an usual Kondo interaction term.

In terms of the electron fields $\psi_{1(2)\sigma}(x)=\frac{1}{\sqrt{2}}
[\psi_{R\sigma}(x)\pm\psi_{L\sigma}(-x)]$, the interaction terms between the
conduction electrons and the local spin can be written as
\begin{eqnarray}
\displaystyle{ \sum_{j}J^{j}_{1}[s_{Rj}(0)+s_{Lj}(0)]\cdot S_{j}}
&=& \displaystyle{ \sum_{j}J^{j}_{1}[s_{1j}(0)+s_{2j}(0)]\cdot S_{j}}
\nonumber \\
\displaystyle{ \sum_{j}J^{j}_{2}[s_{RLj}(0)+s_{LRj}(0)]\cdot S_{j}}
&=& \displaystyle{ \sum_{j}J^{j}_{2}[s_{1j}(0)-s_{j2}(0)]\cdot S_{j}}
\label{26}\end{eqnarray}
where ${\bf s}_{1(2)}(0)=\psi^{\dagger}_{1(2)\alpha}(0)
{\bf \sigma}_{\alpha\beta}\psi_{1(2)\beta}(0)$. We shall treat the system 
by abelian bosonization where the bare potential $J^{z}_{2}$ will be
renormalized to infinity in the low energy region. It is convenient to 
replace the Hamiltonian (\ref{25}) by a renormalized form 
obtained from usual renormalization group method, 
\begin{eqnarray}
H_{eff.} &=& H_{0}+H^{R}_{K} \nonumber \\
H^{R}_{K} &=& \displaystyle{ \sum_{j=x,y}J^{j}_{1}[s_{1j}(0)+s_{2j}(0)]
\cdot S_{j}+\sum_{j=x,y}J^{j}_{2}[s_{1j}(0)-s_{2j}(0)]\cdot S_{j}} 
\label{27} \\
&+& \displaystyle{4\hbar\tilde{v}_{F}\tilde{\delta}[s_{1z}(0)+s_{2z}(0)]S_{z}
+4\hbar v_{F}\delta[s_{1z}(0)-s_{2z}(0)]S_{z}}
\nonumber\end{eqnarray}
where $\tilde{v}_{F}=v_{F}(1-\gamma^{2})^{1/2}$, the bare phase shift
$\tilde{\delta}_{0}=J^{z}_{1}/(4\hbar\tilde{v}_{F})$, and
$\delta=\arctan[e^{(1/g-1)l}\tan(\frac{J^{z}_{2}}{4\hbar v_{F}})]$, where
$l$ is the renormalization scale parameter. 
The renormalization of the bare phase shift $\tilde{\delta}_{0}$ depends on
the electron interaction strength and the phase shift $\delta$. 
At the infrared critical point $\delta^{c}=\pi/2$, the Hamiltonian of the
system has a simple form (see (\ref{30})), $\bar{H}^{c}_{eff.}=
\bar{H}^{c}_{0}+H^{c}_{K}$, where $H^{c}_{K}$ can be rewritten as,
\[
H^{c}_{K}=4\hbar\tilde{v}_{F}\tilde{\delta}[s_{1z}(0)+s_{2z}(0)]S_{z}+
\frac{D}{2\pi\hbar v_{F}}\{J_{2}[e^{-i\Phi_{+s}(0)}S^{+}+h.c.]+
J_{1}[e^{-i\Phi_{+s}(0)-i2\Phi_{-s}(0)}S^{+}+h.c.]\}.  \]
 Using the result\cite{13} $<e^{-i\Phi_{\pm s}(0,t)}e^{i\Phi_{\pm s}(0,0)}>
\sim t^{-1/g}$, as $t\rightarrow \infty$, we can estimate the conformal
dimension of the $\tilde{\delta}$-term in $H^{c}_{K}$.
For the weakly repulsive electron interaction, $1/2<g<1$, 
the $\tilde{\delta}$-term has conformal dimensions $3-1/g$, and is irrelevant,
therefore the renormalized phase shift goes to zero
in the low energy limit (Toulouse limit), $\tilde{\delta}\rightarrow
\tilde{\delta}_{c}=0$. In the case of strongly repulsive electron interaction,
$g\geq 1/2$, the local spin becomes nearly free, 
the $\tilde{\delta}$-term has conformal dimension one, and is marginal, 
the renormalized phase shift therefore takes a finite value in the low
energy limit, $\tilde{\delta}\rightarrow|\tilde{\delta}_{c}|\leq\pi/2$.
We shall take the renormalized phase shift $\tilde{\delta}$ as a
Toulouse parameter, and assume that $\tilde{\delta}\rightarrow
\tilde{\delta}_{c}$ in the Toulouse limit. In the case of $V=0$ 
(no interaction among conduction electrons) and $J^{j}_{2}=0$, the 
system reduces to an usual two-channel Kondo problem, and in the 
Toulouse limit the phase shift takes the value\cite{21,22,23}
$\tilde{\delta}_{c}(V=0)=-\pi/2$. Note that in the abelian bosonization
treatment of the Kondo problem, the spin rotational symmetry is 
artificially broken\cite{21,22,23}. In particular, if an original exchange
interaction has the spin rotational symmetry, the renormalized effective
exchange interaction should retain this symmetry. However, in abelian
bosonization one only renormalize the component $J^{z}_{1(2)}$, and 
the components $J^{x,y}_{1(2)}$ are treated as parameters which represent 
the hybridization between the conduction electrons and the local spin at 
the Toulouse limit $\tilde{\delta}_{c}$. The parameters $J^{x,y}_{1(2)}$ 
should become small at the critical points $\delta^{c}$ 
and $\tilde{\delta}_{c}$
because they are irrelevant. This artificial broken-symmetry originates 
from the bosonic representation of the electron fields, the z-component 
of the spin can be presented as a linear
term of a boson field, then it can be absorbed into the (x,y)-component of the
spin by simple unitary transformations. While under these unitary 
transformations the conformal dimensions of the $J^{x,y}_{1(2)}$-term are
reduced, and they become irrelevant at the critical points $\delta^{c}$ and 
$\tilde{\delta}_{c}$. We believe that
this process should incorporate the renormalization effect of the exchange
interaction potentials $J^{x,y,z}_{1(2)}$ even though one only takes
$\delta^{c}$ and $\tilde{\delta}_{c}$ determined by the renormalized 
$J^{z}_{2}$ and $J^{z}_{1}$, respectively. 
It is well-known\cite{19',21,22} that the abelian bosonization correctly
describes the low energy physical property of the Kondo problem as that of 
Bethe ansatz and conformal field theory treatments\cite{27,27'} 
where the spin rotational symmetry is retained.

Now we derive the effective Hamiltonian at the infrared critical point
$\delta^{c}=\pi/2$ which determines the low energy physical behavior of the 
system. To this purpose, we define an unitary transformation operator which
can be used to treat exactly the $\tilde{\delta}$ and $\delta$ 
terms in (\ref{27}) 
\begin{equation}
\hat{U}^{'}=\exp\{i\frac{2g\tilde{\delta}}{\pi}\Phi_{+s}(0)S_{z}+
i\frac{2\delta}{\pi}\Phi_{-s}(0)S_{z}\}
\label{28}\end{equation}
where $\Phi_{\pm s}(0)=\frac{1}{2}\{[\Phi_{1\uparrow}(0)-\Phi_{1\downarrow}(0)]
\pm [\Phi_{2\uparrow}(0)-\Phi_{2\downarrow}(0)]\}$, $\partial_{x}
\Phi_{1(2)\sigma}(x)=2\pi\rho_{1(2)\sigma}(x)$, and $\rho_{1(2)\sigma}(x)=
\psi^{\dagger}_{1(2)\sigma}(x)\psi_{1(2)\sigma}(x)$. Under the unitary
transformation $\bar{H}^{c}_{eff.}=\displaystyle{ \hat{U}^{'\dagger}
H_{eff.}\hat{U}^{'}}$ and with the gauge transformation
\begin{equation}
\psi_{1\sigma}(x)=\bar{\psi}_{1\sigma}(x)e^{i\theta_{1}}, \;\;\;\;
\psi_{2\sigma}(x)=\bar{\psi}_{2\sigma}(x)e^{i\theta_{2}}, \;\;\;\;
\theta_{1}-\theta_{2}=2\eta\delta S_{z}
\label{29}\end{equation}
where $\eta=+1$ for the $\sigma=\uparrow$ and $\eta=-1$ for $\sigma=
\downarrow$, the Hamiltonian (\ref{27}) becomes at the infrared 
critical point $\delta^{c}=\pi/2$,
\begin{eqnarray}
\bar{H}^{c}_{eff.} &=& \displaystyle{ \hat{U}^{'\dagger}H_{eff.}\hat{U}^{'}=
\bar{H}^{c}_{0}+H^{c}_{K}} \nonumber \\
\bar{H}^{c}_{0}
&=& \displaystyle{ -i\hbar v_{F}\int dx [\bar{\psi}^{\dagger}_{R\sigma}(x)
\partial_{x}\bar{\psi}_{R\sigma}(x)-\bar{\psi}^{\dagger}_{L\sigma}(x)
\partial_{x}\bar{\psi}_{L\sigma}(x)]} \nonumber \\
&+& \displaystyle{ \frac{V}{2}\int dx[\bar{\rho}_{R\sigma}(x)
\bar{\rho}_{R\sigma}(-x)+\bar{\rho}_{L\sigma}(x)\bar{\rho}_{L\sigma}(-x)]}
\nonumber \\
&+& \displaystyle{ \frac{eV_{g}}{2}\sum_{\sigma}\int dx
[\bar{\rho}_{R\sigma}(x)-\bar{\rho}_{L\sigma}(x)]} \label{30} \\
H^{c}_{K} &=& \displaystyle{ \frac{J_{2}D}{2\pi\hbar v_{F}}\{e^{-i(1+
\frac{2g\tilde{\delta}}{\pi})\Phi_{+s}(0)}S^{+}+e^{i(1+
\frac{2g\tilde{\delta}}{\pi})\Phi_{+s}(0)}S^{-}\}} \nonumber \\
&+& \displaystyle{ \frac{J_{1}D}{2\pi\hbar v_{F}}\{e^{-i(1+
\frac{2g\tilde{\delta}}{\pi})\Phi_{+s}(0)}e^{-i2\Phi_{-s}(0)}S^{+}+
e^{i(1+\frac{2g\tilde{\delta}}{\pi})\Phi_{+s}(0)}e^{i2\Phi_{-s}(0)}S^{-}\}}
\nonumber\end{eqnarray}
where $\bar{\psi}_{R\sigma}(x)=[\bar{\psi}_{1\sigma}(x)+\bar{\psi}_{2\sigma}]
/\sqrt{2}$, $\bar{\psi}_{L\sigma}(-x)=[\bar{\psi}_{1\sigma}(x)-
\bar{\psi}_{2\sigma}(x)]/\sqrt{2}$, $\bar{\rho}_{R(L)\sigma}(x)=
\bar{\psi}^{\dagger}_{R(L)\sigma}(x)\bar{\psi}_{R(L)\sigma}(x)$, and $S^{\pm}=
S_{x}\pm iS_{y}$. Here we have taken the transformation (\ref{11}) with
spin indices included 
and $J_{1}+J_{2}=J^{x}_{1}=J^{y}_{1}$ as well as $J_{1}-J_{2}=
J^{x}_{2}=J^{y}_{2}$. The interaction Hamiltonian $H^{c}_{K}$ between the
conduction electrons and the local spin of the quantum dot depends only on
the boson fields $\Phi_{\pm s}(0)$, and cannot be explicitly
presented by the electron
fields $\bar{\psi}_{R(L)\sigma}(x)$. The last term in $H^{c}_{K}$ has a high
conformal dimension $(4+(1+\frac{2g\tilde{\delta}}{\pi})^{2})/(2g)$, and can 
be neglected in lowest order approximation. Therefore, at the infrared 
critical point $\delta^{c}=\pi/2$, only the boson field $\Phi_{+s}(0)$ 
interact with the local spin of the quantum dot. For simplicity, we shall
use a spinless fermion to represent the local spin, $S^{-}=f$,
$S^{+}=f^{\dagger}$, and $S_{z}=f^{\dagger}f-1/2$, the Hamiltonian
$H^{c}_{K}$ can be rewritten as
\begin{equation}
H^{c}_{K}=K[\Psi^{\dagger}(0)f+f^{\dagger}\Psi(0)]
\label{31}\end{equation}
where $K=J_{2}(\frac{D}{2\pi\hbar v_{F}})^{1/2}$, and $\Psi(0)=
(\frac{D}{2\pi\hbar v_{F}})^{1/2}\exp\{-i(1+\frac{2g\tilde{\delta}}{\pi})
\Phi_{+s}(0)\}$ is an anyon field which anticommutates with the fermion field
$f$. 

Using Dyson equation and $H^c_K$, we can obtain the Green function of 
the fermion $f$, $G^{-1}_{f}=G^{-1}_{f0}-\Sigma_{f}$, where the 
self-energy $\Sigma_{f}$ is given by $\Sigma_{f}(t)=K^{2}G_{\Psi}(t)$,
where $G_{\Psi}$ is the Green function of the anyon field 
$\Psi(0)$. With the Hamiltonian $\bar{H}^{c}_{eff.}$ (\ref{30}), we obtain
the correlation function\cite{13}, $<e^{i\Phi_{+s}(0,t)}e^{-i\Phi_{+s}(0,0)}>
\sim t^{-1/g}$ for large $t$. Therefore, the anyon Green function has the
asymptotic behavior, $G_{\Psi}(t)\sim t^{-(1+\frac{2g\tilde{\delta}_{c}}{\pi}
)^{2}/g}$, where $\tilde{\delta}_{c}$ is the value of the Toulouse parameter
$\tilde{\delta}$ in the Toulouse limit, and the fermion Green function
can be written in the low energy limit, 
\begin{equation}
G_{f}(\omega)=\frac{1}{\omega-ib
\omega^{-1+(1+\frac{2g\tilde{\delta}_{c}}{\pi})^{2}/g}}
\label{32}\end{equation}
where $b$ is a
constant proportional to $K^{2}$. Usually, the fixed point $\tilde{
\delta}=0$ is unstable due to the Kondo interaction between the conduction
electrons and the local spin. There exists an infrared critical point 
(corresponding to the Toulouse limit) defined by the renormalized 
$J^{zR}_{1}$ or the Toulouse parameter $\tilde{\delta}_{c}$, the Hamiltonian
at this infrared critical point determines the low energy physical behavior
of the system. In the case of $V=0$ and $J^{j}_{2}=0$, the system reduces
to an usual two-channel Kondo problem, and has an infrared critical point
corresponding to $\tilde{\delta}_{c}=-\pi/2$. In the presence of the
electron-electron interaction and $J^{j}_{2}\neq 0$, we assume the Toulouse
parameter $\tilde{\delta}_{c}$ satisfies\cite{21,22,23}, 
$0\geq\tilde{\delta}_{c}\geq -\pi/2$. In the case of weak repulsive
electron-electron interaction $\kappa=(1+
\frac{2g\tilde{\delta}_{c}}{\pi})^{2}/g<2$, the fermion Green function has the 
asymptotic behavior, $G_{f}(t)\sim t^{-2+\kappa}$ as $t\rightarrow\infty$.
For the strong repulsive electron-electron interaction, i.e., $\kappa\gg 2$,
the fermion Green function $G_f$ is a free-fermion Green's function at
low energy, i.e., the local spin becomes free, which is consistent with 
our previous calculation\cite{13,24}.

Now we consider the transmission and reflection of conduction electrons 
on the local spin of the quantum dot. Under the unitary transformation
$\hat{U}^{'}$, the electron fields $\psi_{R(L)\sigma}(x)$ become
\begin{eqnarray}
\hat{U}^{'\dagger}\psi_{R\sigma}(x)\hat{U}^{'} &=& \displaystyle{
\frac{1}{2}e^{i\eta(g\tilde{\delta}+\delta)S_{z} sgn(x)+i\theta_{1}}\{[1+
e^{-i2\eta\delta S_{z}(1+sgn(x))}]\bar{\psi}^{'}_{R\sigma}(x)} \nonumber \\
&+& \displaystyle{ [1-
e^{-i2\eta\delta S_{z}(1+sgn(x))}]\bar{\psi}^{'}_{L\sigma}(-x)\}} \nonumber \\
\hat{U}^{'\dagger}\psi_{L\sigma}(x)\hat{U}^{'} &=& \displaystyle{
\frac{1}{2}e^{-i\eta(g\tilde{\delta}+\delta)S_{z} sgn(x)+i\theta_{1}}\{[1-
e^{-i2\eta\delta S_{z}(1-sgn(x))}]\bar{\psi}^{'}_{R\sigma}(-x)} \label{33} \\
&+& \displaystyle{ [1+
e^{-i2\eta\delta S_{z}(1-sgn(x))}]\bar{\psi}^{'}_{L\sigma}(x)\}}
\nonumber\end{eqnarray}
where $\bar{\psi}^{'}_{R(L)\sigma}(x)=\exp\{ieV_{g}x\theta(\pm x)
/(\hbar v_{F})\}\bar{\psi}_{R(L)\sigma}(x)$, and $\theta(x)=1$ for
$x>0$ and $\theta(x)=0$ for $x<0$. According to these transformations,
we can obtain the reflection and transmission rates of the electrons,
${\cal R}=\sin^{2}(\delta)$, and ${\cal T}=\cos^{2}(\delta)$, respectively,
which is independent of the Toulouse parameter $\tilde{\delta}$, and is the 
same as that of the barrier scattering (\ref{10}). At the infrared critical
point $\delta^{c}=\pi/2$, the electron transmission rate is zero, and the 
electrons are completely reflected on the quantum dot. This result can be 
easily understood because in the abelian bosonization
the exchange interaction term $J^{z}_{2}[s_{RLz}(0)
+s_{LRz}(0)]S_{z}$ just provides a backward scattering of the conduction 
electrons on the local spin.

With the continuity equation $e\partial_{t}\sum_{\sigma}[\rho_{R\sigma}(x,t)
+\rho_{L\sigma}(x,t)]+\partial_{x}\hat{J}(x,t)=0$, we can derive the current
operator
\begin{equation}
\hat{J}(x,t)=ev_{F}(1-\gamma)\sum_{\sigma}[\rho_{R\sigma}(x,t)-
\rho_{L\sigma}(x,t)]
\label{34}\end{equation}
which is the same as the current operator (\ref{13}) in the case of 
barrier scattering $\epsilon_{0}\gg 0$, except the presence of extra spin
indices. Under the unitary transformation $\hat{U}^{'}$, the density 
operators $\rho_{R\sigma}(x)$ become
\begin{eqnarray}
\hat{U}^{'\dagger}\rho_{R\sigma}(x)\hat{U}^{'} &=& \displaystyle{
\frac{1}{2}\{ \bar{\rho}_{R\sigma}(x)+\bar{\rho}_{L\sigma}(-x)+\cos[\delta
(1+sgn(x))][\bar{\rho}_{R\sigma}(x)-\bar{\rho}_{L\sigma}(-x)]} \nonumber \\
&-& \displaystyle{i2\eta\sin[\delta(1+sgn(x))]S_{z}[
\bar{\psi}^{'\dagger}_{L\sigma}(-x)\bar{\psi}^{'}_{R\sigma}(x)-
\bar{\psi}^{'\dagger}_{R\sigma}(x)\bar{\psi}^{'}_{L\sigma}(-x)]\}} \nonumber \\
\hat{U}^{'\dagger}\rho_{L\sigma}(x)\hat{U}^{'} &=& \displaystyle{
\frac{1}{2}\{\bar{\rho}_{R\sigma}(-x)+\bar{\rho}_{L\sigma}(x)-\cos[\delta
(1-sgn(x))][\bar{\rho}_{R\sigma}(-x)-\bar{\rho}_{L\sigma}(x)]} \label{35} \\
&+& \displaystyle{i2\eta\sin[\delta(1-sgn(x))]S_{z}[
\bar{\psi}^{'\dagger}_{L\sigma}(x)\bar{\psi}^{'}_{R\sigma}(-x)-
\bar{\psi}^{'\dagger}_{R\sigma}(-x)\bar{\psi}^{'}_{L\sigma}(x)]\}}
\nonumber\end{eqnarray}
and the current operator $\hat{\cal J}(x,t)=\hat{U}^{'\dagger}\hat{J}(x,t)
\hat{U}^{'}$ is ($x>0$)
\begin{eqnarray}
\hat{\cal J}(x,t) &=& \displaystyle{ ev_{F}(1-\gamma)\sum_{\sigma}
[\alpha\bar{\rho}_{R\sigma}(x,t)+\beta\bar{\rho}_{L\sigma}(-x,t)-
\bar{\rho}_{L\sigma}(x,t)]} \nonumber \\
&-& \displaystyle{ i2ev_{F}(1-\gamma)\sin(2\delta)[s_{LRz}(-x,x)-
s_{RLz}(x,-x)]S_{z}}
\label{36}\end{eqnarray}
where $s_{RL(LR)z}(x,x')=\frac{1}{2}\bar{\psi}^{'\dagger}_{R(L)\alpha}(x)
\sigma^{z}_{\alpha\beta}\bar{\psi}^{'}_{L(R)\beta}(x')$ are non-local spin
operators. Using the Hamiltonian $\bar{H}^{c}_{eff.}$ (\ref{30}), we
obtain the current of the system 
\begin{equation}
{\cal I}=<\hat{\cal J}(x,t)>=\frac{ge^{2}}{\pi\hbar}V_{g}{\cal T}
\label{37}\end{equation}
where the electron transmission rate ${\cal T}$ is given by Eq.(\ref{16}) 
with the bare $\delta_{0}$ replaced by $J^{z}/(4\hbar v_{F})$. The last 
term in (\ref{36}) does not contribute to the current, because at the 
infrared critical point $\delta^{c}=\pi/2$ the right- and left-moving 
electrons are completely separated, so the average of the non-local 
spin operator is zero, $<s_{RL(LR)z}(x,x')>=0$. 

   Note that as in previous cases, to study the low energy properties
of the system with $V_g\neq0$, we need to add a perturbative term 
$\Delta H=-4\hbar v_{F}(\delta^{c}-\delta)[s_{RLz}(0)+s_{LRz}(0)]S_{z}$ 
to the critical Hamiltonian $\bar{H}^{c}_{eff.}$(\ref{30}), where
$\Delta\delta=\delta^{c}-\delta=\arcsin[\cos(\delta)]\ll 1$ presents the 
deviation of the system away from the infrared critical point
$\delta^{c}=\pi/2$. The first order correction of the current from the
perturbation $\Delta H$ is $-4ev_{F}(1-\gamma)\sin(2\delta)
\int^{L/2}_{0}dx\int dt'<[s_{LRz}(-x,x,t)-s_{RLz}(x,-x,t)]S_{z}(t)
\Delta H(t')>/L$, which goes to zero in the infinite quantum wire limit 
$L\rightarrow\infty$, and is proportional to $V_{g}^{4/g-3}$ for the 
finite quantum wire $L\ll \hbar v_{F}/(eV_{g})$. Therefore, the 
correction of the current by the $\Delta H$ can be neglected, and the 
current of the system is presented by ${\cal I}$(\ref{37}) to lowest
order in $V_g$. This result means that the tunneling current through 
the quantum dot has the same low energy behavior for both the cases 
of $\epsilon_{0}>> \Gamma$ ($<d^{\dagger}_{\sigma}(t)d_{\sigma}
(t)>\sim0$) and $\epsilon_{0}<< \Gamma$ ($<d^{\dagger}_{\sigma}(t)
d_{\sigma}(t)>\sim1$). However, in the latter case the spin 
susceptibility of the quantum dot is not zero.

   The tunneling current (\ref{37}) is one of our central results. This 
result is consistent with that of Ref.\cite{12} where the authors used 
open boundary condition, and the magnetic impurity is residing at the 
end of a half-infinite quantum wire. Although similar expression of the 
low energy tunneling current is obtained in both cases, the microscopic
physics appears to be rather different. At present case, the
backward-scattering-like Kondo interaction $J^{i}_{2}[s_{RLi}(0)+
s_{LRi}(0)]\cdot S_{i}$ is relevant for
repulsive electron-electron interaction, and induces the complete 
reflection of electrons on the quantum dot at the critical fixed point
$\delta^{c}=\pi/2$. This relevant interaction also affects the usual 
Kondo interaction term $J^{i}_{1}[s_{Ri}(0)+s_{Li}(0)]
\cdot S_{i}$. On the other hand, in the open boundary case the 
complete reflection of electrons on the magnetic impurity is a 
boundary constraint. The backward-scattering-like Kondo interaction 
term is irrelevant, and contributes only a high order correction to 
the usual Kondo interaction. As far as electronic transport is 
concerned, the tunneling current of the system is determined only by 
the infrared critical fixed point ($\delta^{c}=\pi/2$) Hamiltonian, 
therefore similar tunneling current expression are obtained in these 
two cases. However, the impurity susceptibility is determined by both 
the infrared critical fixed point Hamiltonian as well as the coupling 
between the magnetic impurity and the conduction electrons at 
(or near) the infrared critical point. At present case, the usual 
two-channel Kondo fixed point is unstable due to the 
backward-scattering-like Kondo interaction, while in the open boundary 
case, the two-channel Kondo fixed point is stable for
strong repulsive electron-electron interaction in which the 
backward-scattering-like Kondo interaction term is irrelevant. Therefore,
the impurity susceptibility shows different low energy behavior in
these two cases\cite{12,13,24}. 

\section{ Discussion and Conclusion}

   It will be possible in future experiments to fabricate and study a 
system consist of a quantum dot or an artificial atom connected with an 
infinite quantum wire. The quantum dot or artificial atom can be made very 
small by modulating gate voltages, so that it has well-defined discrete 
energy levels. If the gap between two nearest levels is large compared 
with the hybridization $\Gamma$ between the dot and the quantum 
wire, the low energy physical behavior of the system is controlled by the 
level $\epsilon_{0}$ closest to the Fermi level of the quantum wire. 
With this simple system, one can experimentally study the physical
property of an Anderson model with and without the electron-electron
interaction in the mixed valence and single occupied states.

This system can be presented by an single-impurity Anderson model with
electron-electron interaction, if we assume that the other levels below 
the level $\epsilon_{0}$ are occupied by even electrons, and above it 
are unoccupied. However, this simple model can not be exactly solved 
and the nature of the ground state depends strongly on $\epsilon_{0}$. 
The Hilbert space of the local electron consists of 
$\{|0>, \; |\uparrow>, \; |\downarrow>, \; |\uparrow\downarrow>\}$. In 
the case of $\epsilon_{0}> 0$, the ground state of the local electron 
is essentially just $|0>$. $<d^{\dagger}_{\sigma}(t)d_{\sigma}(t)>
\sim0$, and the quantum dot problem reduces to the problem of simple
barrier-like potential scattering for conduction electrons.
In the case of $\epsilon_{0}\sim0$, the local electron is in
a mixed valence state with $<d^{\dagger}_{\sigma}(t)d_{\sigma}(t)>\;< 1$, 
the local electron orbital is in resonance with the conduction electrons. 
In the case of $\epsilon_{0}<< -\Gamma$, the system is in the Kondo regime. 
The local electron state is singly occupied state with
$<d^{\dagger}_{\sigma}(t)d_{\sigma}(t)>=1$, and the local electron has 
a Kondo exchange interaction with the conduction electrons, i.e., the
system reduces to an one-dimensional Kondo problem with 
electron-electron interaction. In these three cases, the local electron 
orbital has very different states, and the system has very different low 
energy behavior.

   With the bosonization method and phase shift representation, we studied in
this paper the low energy transport behaviors of the system in these 
three limit cases at zero temperature and with a small external 
voltage $V_{g}$. We have demonstrated that the current of
the system has the same low energy power-law behavior 
$\sim V_{g}^{2/g-1}$ for both cases of $\epsilon_{0}>> \Gamma$ and
$\epsilon_{0}<< -\Gamma$, even though the local electrons have completely
different effects on conduction electrons in these two cases. In the 
former case, the local electron orbital only provides a barrier-like 
potential scattering, whereas in the latter, the local electron has
the Kondo-type exchange interaction with the conduction electrons. The 
same low energy current expression originates from the fact that there 
exist similar backward scattering potentials in both cases, $U_{2k_{F}}[
\psi^{\dagger}_{R\sigma}(0)\psi_{L\sigma}(0)+\psi^{\dagger}_{L\sigma}(0)
\psi_{R\sigma}(0)]$, and $J^{z}_{2}[s_{RL}(0)+s_{LR}(0)]S_{z}$, respectively.
Note that in general the low energy properties of the system are different
in these two limit cases, such as the spin susceptibility of the
local electron. Generally, in the case of $\epsilon_{0}<< -\Gamma$, the 
local electron spin susceptibility shows power-law behavior at low
energy, but for strong enough repulsive electron-electron interaction, 
the local spin becomes nearly free. This prominent property of 
one-dimensional Kondo problem originates from the almost complete 
reflection of the electrons on the local spin in the low energy limit. 
As a result the electron density of states near the local spin becomes 
small, and the magnitude of the spin exchange terms 
$s^{+}_{R(L)}(0)S^{-}$ and $s^{-}_{R(L)}(0)S^{+}$ are strongly reduced. 
In the case of $\epsilon_{0}\sim 0$, the local electron is in the mixed
valence state. The system shows different low energy behavior from that 
of the above two cases. we demonstrate that even though the electrons 
are almost completely reflected by the local electron in the low energy 
region, the resonance between the local and conduction electrons enhances 
the tunneling current through the quantum dot. In this resonance case, 
the system has the same low energy transport properties as that of a 
simple model of a resonant level coupling to two half-infinite 
quantum wires where electron fields satisfies open boundary conditions
at the resonant level site, and the tunneling current is proportional to
$V_{g}$. The advantage of the phase shift representation is that by using 
the simple unitary transformation we can directly obtain the tunneling current
without calculating correlation functions.

We acknowledge support of HKRGC through Grant No. UST6143/97P.

\end{document}